\documentstyle[12pt]{article}
\font\gotb eufm10 scaled \magstep2
\newcommand{\bb}{\bibitem}
\newcommand{\cc}{\cite}
\newcommand{\lll}{\lambda}
\newcommand{\lt}{\left}
\newcommand{\rt}{\right}
\newcommand{\vp}{\varphi}
\newcommand{\R}{\hat R}
\newcommand{\Q}{\hat Q}
\newcommand{\A}{\hat A}
\newcommand{\B}{\hat B}
\newcommand{\SS}{\hat S}
\newcommand{\PP}{\hat P}
\newcommand{\AAA}{\hbox{\gotb A}}
\newcommand{\beq}{\begin{equation} \label}
\newcommand{\bea}{\begin{eqnarray} \label}
\newcommand{\eeq}{\end{equation}}
\newcommand{\eea}{\end{eqnarray}}
\newcommand{\nn}{\\ \nonumber}
\newcommand{\rr}[1]{(\ref{#1})}
\author{D.A.Slavnov\thanks{ E-mail: slavnov@theory2.npi.msu.su }} 

\title{ Quantum mechanics with the permitted hidden parameters}

\date{ Department of Physics, Moscow State University, Moscow 119899, Russia. }

\begin{document}

\maketitle \ \\ [-1cm]

\begin{abstract}

Within the framework of the algebraic approach the problem of hidden parameters in quantum mechanics is surveyed. It is shown that the algebraic formulation of quantum mechanics permits introduction of a specific hidden parameter, which has the form of nonlinear functional on the algebra of observables. It is found out that the reasoning of von Neumann and of Bell about incompatibility of quantum mechanics with hidden parameters is inapplicable to the present case. 
\end{abstract}

During many years in interpretation of quantum mechanics the leading position occupies the approach which goes back to the Founding Fathers of quantum mechanics: Bohr, Heisenberg, Dirac and others. Frequently this approach is named orthodox direction. Under the present circumstances I would name it the standart one.

At early stages of development of quantum mechanics this approach called forth many objections. The Einstein criticism of it is the most known. First of all it is very difficult to agree that the principle of causality (determinism) is substituted in the orthodox approach by a principle of probabilistic character of the majority of physical processes. There are well-known words of Einstein about the God playing dice.

From the point of view of usual human logic a model "with hidden parameters" (see Bohm~\cc{bom}) seems the much more acceptable. However, after the works of von Neumann~\cc{von} and Bell~\cc{bel} it was became firmly established opinion that models with hidden parameters is a false direction in the development of quantum mechanics. In the proposed work I want to rehabilitate models with hidden parameters. I think, I could construct a model, for which the arguments of von Neumann and Bell would turn out inapplicable.

In the famous monography~\cc{von} von Neumann asserts that the models with hidden parameters are incompatible with the basic theses of quantum mechanics and without a radical modifications of these theses hidden parameters cannot be introduced.

As basic theses of quantum mechanics von Neumann accepts, in particular, such postulates:

N.1. To each observable quantity $ \A $ there corresponds a linear operator $ {\cal A} $ in some Hilbert space.

N.2. If the operators $ {\cal A} $ and $ {\cal B} $ correspond to observable quantities $ \A $ and $ \B $, the operator $ {\cal A} + {\cal B} $ corresponds to the quantity $ \A + \B $ (simultaneous measurability of quantities $ \A $ and $ \B $ is not necessary).

I think these postulates are too restrictive. They can be weakened without any breaking of the mathematical formalism of quantum mechanics. The idea that a Hilbert space and the linear operators in it should not be primary elements of quantum theory is not new. It was this idea that became the basis of development of the algebraic approach in quantum field theory~\cc{emch}.

Following this direction, we shall assume that the elements of some algebra correspond to observable quantities. Since all observable quantities are real, the experiment testifies in favour of a real algebra. It is quite possible, the Jordan algebra of observables can set up a claim to this role (see~\cc{emch}). However, in this algebra the operation of multiplication is defined in a rather complicated manner, and there is no property of associativity. In this connection it seems to me justified to leave the framework of directly observable quantities and to consider their complex combinations, which further will be referred to as dynamical quantities.

Correspondingly we shall accept a postulate that the elements of an involutive associative noncommutative algebra~\AAA{} correspond to dynamical quantities, obeying:\\
1) for each element $\R\in\AAA$ there is the hermitian element $\A$ $(\A^*=\A)$ that $\R^*\R=\A^2$,\\
2) if $\R^*\R=0$, then $\R=0$.

The hermitian elements of algebra~\AAA{} correspond to observable quantities. They are a latent form of the observable quantity. The explicit form of the observable should be some number. In other words, for the representation of the explicit form of observables on hermitian elements of algebra~\AAA{} some functional $\vp$ should be set: $\vp(\A)=A$ -- real number.

The physically, the latent form of an observable $\A$ becomes explicit as a result measurements. It means, that the functional $\vp(\A)$ gives the value of the observable $\A$, which could obtain in a {\it concrete (individual)} measurement. This physical interpretation of the functional ~$\vp(\;)$ can be taken as its definition.

Now we shall undertake a decisive step in the direction of introducing the principle of causality in quantum theory. We postulate that the functional~$\vp(\;)$ is determined by structure of the considered quantum object and it do not depend on the information, which the observer has.

The proposed further structure seems to me sufficiently probable. At the same time, all further deductions do not depend on particular details of this structure.

All modern development of quantum theory testifies in favour of locality of this theory. Besides this, the measuring device responds to an elementary quantum object as the whole (corpuscular character of quantum object). Therefore, most probably, in an elementary quantum object there is a local component which is the carrier of corpuscular and dynamical properties (of dynamical degrees of freedom) of the object. It gives a reason to suppose that the dynamical quantities $\A$ are related to this local component of the quantum object.

On the other hand, the quantum object has wave properties. It means that some wave field is related to the quantum object. Firstly, the wave field is a nonlocal formation; secondly, it is a system with infinite number of degrees of freedom. That is, the quantum object should have some nonlocal component, which is not the carrier of dynamical degrees of freedom. This component can be the carrier of wave (phase) degrees of freedom.

If we suppose that the observable $\A$ is not the source of wave properties, then the unique source of these properties is the functional $\vp(\;)$. Therefore, it is natural to assume that this functional is determined by a nonlocal component of the quantum object. We shall call the structure of this nonlocal component the {\it physical state} of the quantum object. Let us believe that the physical state {\it uniquely} determines the reaction of the measuring device to the concrete quantum object. This supposition corresponds to the basic idea of the so-called PIV-model (see~\cc {hom}).

Let us note that in the standart quantum mechanics the state  is a  mathematical object, an element of Hilbert space, only. Similarly, in the algebraic approach the state is an element of space which is dual to algebra~\AAA{} (the set of linear functionals on algebra~\AAA). In the proposed approach the physical state is determined by the material structure of the quantum object. At the same time the physical state is associated with mathematical notion --- with a functional (nonlinear) on algebra~$\AAA$.

Now we shall introduce a construction, which corresponds to pure state in the standart quantum mechanics. Just as in the standart quantum mechanics we shall suppose that only the mutually commuting observables are simultaneously measurable. Let $\{\Q\}$ be some maximal set of the mutually commuting hermitian elements of the algebra~\AAA. The functional $\vp(\;)$ maps the set $\{\Q\}$ into the set of real numbers $$
\{\Q \}\stackrel{\vp}{\longrightarrow}\{Q=\vp(\Q)\}.
$$
For different functionals $\vp_i(\;)$, $\vp_j(\;)$ the sets $\{\vp_i(\Q)\}$, $\{\vp_j(\Q)\}$ can either differ or coincide. If for all $\Q \in \{\Q\}$ is valid $\vp_i(\Q)=\vp_j(\Q)=Q$, then we shall term physical functionals $\vp_i(\;)$ and $\vp_j(\;)$ as $\{Q\}$-equivalent functionals. Let $\{\vp\}_Q$ be the set of all physical states, with which there correspond $\{Q\}$-equivalent functionals. Let us call the set $\{\vp\}_Q$ a quantum state, and we shall designate it~$\Psi_Q$.

Let us make one fundamental supposition about the physical state. It does not follow from the experiment directly. But, on the one hand, it allows to reproduce the main statements of the standart quantum mechanics. On the other hand, within the framework of the proposed method, if one drops this supposition, it is possible to obtain a series of deductions (in particular, Bell inequality), contradicting the experiment.

So, let us assume that each physical state is unique, i.e. in the nature there are no two identical states, and the physical states never repeat. It is possible to suppose that the physical state is determined by all the previous history of a concrete physical object and this history for each object is individual. In particular, in two different experiments we necessarily deal with two different physical states. It follows from this supposition that each physical state $ \vp $ can manifest itself as the functional $\vp(\;)$ no more than in one experiment.

It also follows from this supposition that the physical state cannot be uniquely fixed. Really, to fix a functional $ \vp (\;) $, we should know its value $ \vp (\B) $ for all independent observables $ \B $. Physically it is not realizable. In one experiment we can find $ \vp (\B_i) $ only for mutually commuting observables $ \B_i $, and in different experiments we necessarily deal with different functionals $\vp(\;)$.

The most strict fixing of the physical state $\vp$, which can be realized practically, consists in assigning it to some set $\{\vp\}_Q$, i.e. to a certain quantum state. For this purpose it is sufficient to perform measurements of mutually commuting observables. It is possible to restrict oneself to independent measurements only. In principle it can be done in one experiment. Thus, in essence we can not have the complete information about a physical state of a concrete physical object.

The incompleteness of the information about a physical state of a quantum object allows us to do only probabilistic predictions for the further behaviour of this object. Thus, the originator of probabilistic character of quantum mechanics is not the God who is playing dice, but quite material reasons. There are two reasons: uniqueness of each physical state and the impossibility to measure all observables in one experiment (in one physical state).

It follows from the uniqueness of physical state that the value of each observable can become explicit not in any physical state. So, if the value of the observable $\A$ is measured in a certain physical state $\vp$, the value of the observable $\B$, noncommuting with the observable $\A$, cannot be measured in the state $\vp$. In this sense the state $\vp$ is irrelevant for the observable~$\B$.

 Correspondingly for each observable $\A$ it is expedient to introduce a notion --- the set $[\vp]^{\A}$ of relevant (actual) states. This set contains all physical states, in which the observable $ \A $ was ever measured. This physical interpretation of the set $[\vp]^{\A}$ can be taken as its starting definition. However, with the help of the functionals $\vp(\;)$, which correspond to the physical states $ \vp\in [\vp]^{\A}$, we want to describe the results not only really performed measurements. We want also to describe the observed dates, which could be performed in the past or the future. Therefore, it is necessary to expand the initial definition and to suppose that the set $[\vp]^{\A}$ is determined by all physically admissible measurements of the observable~$\A$.

Many measurements are incompatible. Including certain measurements in the set of admissible ones, we should exclude others. Thus, there are many variants of expansion of the initial definition of the set $[\vp]^{\A}$.  What from these variants corresponds to the real physical situation, depends on the choice of the measuring device. Here the observer realizes the freewill. The sets of relevant states for noncommuting observables should have empty intersection.

For the observable $\A$ the set of the $\{Q\}$-equivalent relevant states will be designated~$\{\vp\}^{\A}_Q$. The notion, corresponding to this symbol, is not internally inconsistent. Really, let us consider evolution of a quantum object (its physical state) in the time interval, when the interaction with classical measuring device is absent. Let us suppose that the evolution of the physical state of the quantum object is controlled by the equation of motion, which is coordinated with the usual quantum-mechanical equation of motion.

 Namely we shall assume that the corresponding functional $\vp(\A)$ varies as follows: 
\beq {15}
\vp_0(\A)\to\vp_t(\A) \equiv\vp_0(\A(t)),
\eeq
where
\beq{16}
\frac{d\,\A(t)}{d\,t}=\frac{i}{\hbar}\Big[\hat H, \A(t)\Big],\qquad \A(0)=\A.
\eeq
Here $\hat H$ ($\hat H\in\AAA$) is the Hamiltonian of the quantum object.

 The equations~\rr{15} and~\rr{16} quite uniquely describe temporal evolution of the physical state. Other matter, that with the help of the experiment we can determine the initial value of the functional $\vp_0(\A)$ only for elements, belonging to some set $\{\vp\}_Q$ of mutually commuting elements.

Let it be known that at the instant of time $t=0 $ the physical state $\vp$ of the quantum object belongs to the quantum state $\{\vp\}_Q$. From the equations \rr{15},~\rr{16} follows that at the instant $t$ the physical state of this object will belong to the quantum state $\{\vp\}_{Q(t)}$. Thus, making the choice of the physical states $\vp\in\{\vp\}_Q$ with the help of measurements at the instant of time $t=0 $, we shall obtain with guarantee a physical state $\vp_t\in\{\vp\}_{Q(t)}$ at the instant $t$. Further we can produce measurement of any observable $\A\in\AAA$.

Now we shall fix properties of the functional $\vp$. We shall require that on the hermitian elements of algebra $\AAA${} the functional $\vp(\;)$, corresponding to the relevant states, should satisfy the following postulates: 
\bea{4}
 \quad &1)&\vp(\lll \hat I)=\lll, \ \hat I\mbox{ is the unity of algebra } \AAA, \ \lll \mbox{ is a real number}; \nn {} & 2)& \vp(\A+\B)=\vp(\A)+\vp (\B), \ \vp(\A\B)=\vp(\A)\vp(\B), \mbox{ if } [\A,\B]=0; \nn {} &3)&\vp(\A^2)\ge 0; \qquad \sup_\vp\vp(\A^2)=0, \mbox{ iff }\A=0; \nn{} & 4)& \mbox{for each set } \{Q \} \mbox{ and each } \A \nn {} &\quad&\mbox{there is } \lim_{n\to\infty} \frac{1}{n} \sum^n_{i=1}\vp_i(\A) \equiv\Psi_Q(\A),\nn{}&\quad&\mbox{where } \{\vp_1,\dots,\vp_n \} \mbox{ is a random sample of the set } \{\vp\}_Q^{\A};\nn{}& 5)& \mbox{for every } \A, \; \B \quad \Psi_Q(\A+\B)=\Psi_Q(\A)+\Psi_Q(\B).
\eea

Let us note that these postulates are not arbitrary requirements. In the mathematical form they fix properties, which are the results of realistic measurements of observables. The functional $\Psi_Q(\;)$, appearing in the fourth postulate, has the meaning of the functional $\vp(\;)$, which is averaged over all $\{Q\}$-equivalent relevant states. Symbolically it can be represented in the form of Monte-Carlo integral over the relevant states: 
$$
\Psi_Q(\A)=\int_{\vp\in \{\vp\}_Q^{\A}} d\mu(\vp)\, \vp(\A).
$$
We shall note that
$$
\int_{\vp\in\{\vp\}_Q^{\A}} d\mu(\vp)=1.
$$
Let us assign the functional $\Psi_Q(\;)$ to each quantum state $\Psi_Q$. The fourth postulate assumes that this functional does not depend on a concrete random sample. Further we shall use the term "a quantum state $\Psi_Q$" for both the set $\{\vp\}_Q$ of physical states and for the corresponding functional $\Psi_Q(\;)$ (the quantum average).

 Any element $\R$ of the algebra $\AAA${} can be represented in the form $\R=\A+i\B$, where $\A$ and $\B$ are hermitian elements. Let us extend the functional $\Psi_Q(\;)$ onto the elements $\R$ with the help of equality $\Psi_Q(\R)=\Psi_Q(\A)+i\Psi_Q(\B)$. 
 
Let $\R^*\R =\A^2\in\{\Q\}$ ($\A^* =\A$). If $\vp\in\{\vp\}_Q$, then $\vp(\R^*\R)=A^2$. Therefore, 
 $$
 \lt.\Psi_Q(\R^*\R)=A^2=\vp(\A^2)\rt|_{\vp\in\{\vp\}_Q}. 
$$
 From here 
$$
 \|\R\|^2 \equiv\sup_Q\Psi_Q(\R^*\R)= \sup_{\vp\in[\vp]^{\A}}\vp(\A^2)>0, \mbox{ if} \R\ne0.  
$$

Since $\Psi_Q(\;)$ is a positive linear functional, the Cauchy-Bunkyakovsky-Schwarz  inequality is valid 
 $$
\Psi_Q(\R^*\SS)\Psi_Q(\SS^*\R) \le \Psi_Q (\R^*\R) \Psi_Q (\SS^*\SS).
 $$
Therefore, (see \cc{emch}) for $\|\R\|$ the postulates of norm of the element $\R$ are fulfilled: $\|\R+\SS\|\le \|\R\|+\|\SS\|$, $\|\lll\R\|=|\lll|\,\|\R\|$, $\|\R^*\|=\|\R\|$.
The algebra~\AAA{} become to a Banach space after supplement over norm $\|\;\|$. Since $\vp([\A^2]^2)=[\vp(\A^2)]^2$, then $\|\R^*\R\| =\|\R\|^2$, i.e. the algebra~\AAA{} is $C^*$-algebra.

Thus, from postulates~\rr{4} follows that $\Psi_Q(\;)$ is a positive linear functional on a $C^*$-algebra, satisfying to the normalization condition $\Psi_Q(I)=1$. Therefore, according to the  Gelfand-Naumark-Segal construction (see \cc {emch}), the functional $\Psi_Q(\;)$ canonically generates the representation of the algebra~\AAA{} by linear operators in some Hilbert space. In other words, in the proposed approach it is possible  to reproduce the mathematical formalism of the standart quantum mechanics completely.

At the same time, as opposed to the standart quantum mechanics, only the mathematical notions --- vectors of Hilbert space (wave functions) and the operators in Hilbert space --- are not the primary elements of the theory. They arise only at the second stage. In the proposed approach the primary elements are the observables and physical states, which are related directly to the material structure of the quantum object and they do not depend on the observer.

It is necessary to emphasize that the quantum states $\Psi_Q(\;)$ have some subjective element. The matter is that one physical state can belong to different quantum states $\{\vp\}_Q$ and $\{\vp\}_P$. That is, $\vp\in\{\vp\}_Q\cap \{\vp\}_P$, where the states $\{\vp\}_Q$ are classified by values of the set $\{\Q\}$ of mutually commuting observables $\Q$, and $\{\vp\}_P$ are classified by values of observables $\PP\in\{\PP\}$. The observables $\Q$ and $\PP$ do not commute among themselves. Then depending on what set ($\{\Q\}$ or $\{\PP\}$) we shall choose for the classification, the physical state $\vp$ will be referred either to the quantum state $\{\vp\}_Q$ or to the quantum state $\{\vp\}_P$.

This subjective component in quantum state is the origin of the famous Einstein-Podolsky-Rosen paradox~\cc{epr}. The variant of experiment, proposed Bohm~\cc{bom} for demonstrating the paradox, looks as follows.

Let a spin-zero particle decay into two particles $A$ and $B$ with spins 1/2 which scatter at large distance. Let us measure a projection of spin onto the axis $z$ for the particle $A$. Let the result will be $S_z(A)$. Then, using the conservation law, we can state that for the particle $B$ the projection of spin onto the axis $z$ is equal $S_z(B)= -S_z(A)$ with absolute probability. It means that the quantum state of the particle $B$ corresponds to such value of the projection of spin onto axis $z$. However, for the particle $A$ we could measure the projection of spin onto axis $x$. Let the result be $S_x(A)$. Then we could state that the particle $B$ is in the quantum state, which corresponds to the value $S_x(B)=-S_x(A)$ of the observable $\hat S_x(B)$. Thus, by his wish the observer "drives" the particle $B$ either in the quantum state $\{\vp \}_{-S_z(A)}$, or in the quantum state $\{\vp\}_{-S_x(A)}$, without acting on the particle $B$ physically in any way.

In the proposed approach the paradox does not arise. In both cases the physical state (the objective reality) will be same $\vp\in\{\vp\}_{-S_z(A)}\cap\{\vp\}_{-S_x(A)}$. The various quantum states of the particle $B$ arise due to a subjective choice by the observer of a device for measurement of the observable $\A$.

Such interpretation of the results of the proposed experiment is the adequate answer onto the problem posed in paper~\cc{epr}. It is known that the explanation, which for this paradox was given by Bohr, did not satisfy Einstein.

It is seen from this example that the physical state~$\vp$ can be considered as a specific hidden parameter. On the one hand, a particular physical state~$\vp$ corresponds to a concrete event. In this sense it is a parameter. On the other hand, the physical state cannot be fixed uniquely with the help of experiment; it is possible only to establish its membership in this or that ensemble (quantum state). In this sense~$\vp$ is a hidden parameter.

Von Neumann has made a deduction about the impossibility of existence of hidden parameters from the statement that any quantum-mechanical ensemble has dispersion. That is, in this ensemble not all average squares of observables are equal to squares of average corresponding quantities. In the monography~\cc{von} it is proved that such ensemble cannot be decomposed into zero-dispersion subensembles, as they simply do not exist. This statement appears as a consequence of postulates~N.1 and~N.2.

These postulates are valid for the quantum state, considered in the present work. However, von Neumann's reasoning is invalid for a physical state. A physical state is zero-dispersion "subensemble" comprising one element. Its existence is not forbidden since "average" values of observables over such ensemble are defined by the {\it nonlinear} functional $\vp(\;)$.

By virtue of nonlinearity of the functional $\vp(\;)$ the physical state is not related to any representation of algebra of observables. Therefore, on physical states the observables are not representable by linear operators on a Hilbert space and the postulates N.1 and N.2 are not valid for them.

In fact, in the monography~\cc{von} it is shown that the linearity of a state is in the conflict with  causality and the hypothesis about the hidden parameters. Herefrom von Neumann has made a deduction that causality is absent at the microscopic level, and the causality occurs due to averaging over large number of noncausal events at the macroscopic level.

In the present work, the same conflict is proposed to solve in the opposite way. It is suggested to consider that there is causality at the level of a single microscopic phenomenon but linearity is absent. The linearity of the (quantum) state occurs due to averaging over quantum ensemble. The transition from the single phenomenon to quantum ensemble replaces the initial determinism by probabilistic interpretation.

Besides von Neumann's reasoning, there is one more argument, not less famous, against schemes with hidden parameters, it is Bell inequality~\cc{bel}. We shall reproduce a typical deduction of this inequality.

Let a quantum object $Q$ (particle with spin 0 in the elementary variant of experiment) decay into two objects $A$ and $B$ (particles with spins 1/2). The objects $A$ and $B$ scatter at large distance and hit detectors $D(A)$ and $D(B)$, respectively, in which the measurements are independent. The object $A$ has a set of observables $\A_a$ (double projection of spin onto the direction $a$). The observables corresponding to different values of index $a$ are not simultaneously measurable. Each of the observables can take two values $\pm1$. In a concrete experiment the device $D(A)$ measures the observable $\A_a$ with a particular index $a$. For the object $B$ everything is similar.

Let us assume that the quantum object $Q$ has a hidden parameter $\lll$. In each individual event the parameter $\lll$ has a particular value. The distribution of events over the parameter $\lll$ is characterized by a measure $\mu(\lll)$ with the usual properties $\mu(\lll)\ge0$, $\int d\mu(\lll)=1$. All the magnitudes, connected with an individual event, depend on the parameter $\lll$. In particular, the values of observables $\A_a$ and $\B_b$, obtained in a concrete experiment, are functions $A_a(\lll)$, $B_b(\lll)$ of the parameter $\lll$.

For individual event the correlation of observables $\A_a$ and $\B_b$ is characterized by the magnitude $A_a(\lll)B_b(\lll)$. The average value of the magnitude is referred to as correlation function $E(a,b)$: $$
E(a,b)=\int d\mu(\lll)\, A_a(\lll) \, B_b(\lll).
$$

Giving various values to the indexes $a$ and $b$ and taking into account that
 \beq{21}
 A_a(\lll)=\pm1, \quad B_b(\lll)=\pm1,
 \eeq
we shall obtain the following inequality
 \beq {22}
|E(a,b)-E(a,b')|+|E(a',b)+E(a',b')|\le
 \eeq
$$
\le\int d\mu(\lll)\,[|A_a(\lll)|\,|B_b(\lll)-B_{b'}(\lll)|+ |A_{a'}(\lll)| \, |B_b(\lll)+B_{b'}(\lll)|] = $$
 $$
=\int d\mu(\lll)\, [|B_b(\lll)-B_{b'}(\lll)|+|B_b(\lll)+B_{b'}(\lll)|]. $$
In the right-hand side of the formula \rr{22}, due to equalities \rr{21}, one of the expressions
 \beq{23}
|B_b(\lll)-B_{b'}(\lll)|, \qquad |B_b(\lll)+ B_{b'}(\lll)|
 \eeq
is equal to zero, and the other is equal to two for each value of $\lll$. From here we obtain the Bell inequality 
\beq{24}
|E(a,b)-E(a,b')|+|E(a',b)+E(a',b')| \le 2.
\eeq

The correlation function $E(a,b)$ is easily calculated within the framework of the standard quantum mechanics. In particular, when $A$ and $B$ are particles with spin 1/2 
\beq{25}
E(a,b)=-\cos\theta_{ab}, \qquad \theta_{ab}\mbox{ is an angle between } a\mbox{ and }b.
\eeq
It is easy to verify that there are directions $a,b,a',b'$, for which formulas~\rr{24} and~\rr{25} contradict each other.

 It looks as if it were possible to repeat this derivation in the model, proposed here, having made replacements of type $A(\lll)\to\vp(\A)$, $B(\lll)\to\vp(\B)$, $\int d\mu(\lll)\dots\to \int_{\vp\in \{\vp \}_Q^{\A\B}} d\mu(\vp)\vp (\dots)$. However, this opinion is erroneous. It is essential for derivation of the Bell inequality that in the left-hand side of the inequality~\rr{22} it is possible to collect all terms in one integral over the parameter~$ \lll $. It is not valid for the quantum average, which substitutes this integral, since it is necessary to integrate over relevant states in it. The elements, appearing in different correlation functions, $\A_a\B_b $, $\A_{a'}\B_b$, $\A_a\B_{b'}$, $\A_{a '}\B_{b'}$, do not commute among themselves. Therefore, sets of relevant states, corresponding to these operators, do not intersect. In derivation of the inequality~\rr{24} we tacitly supposed that expression~\rr{23} exist for each $\lll$. However, there is no physical state $\vp$, which would be a relevant state for both the observable $\B_b$ and the observable $\B_{b'}$.

In the forecited discussion of the Bell inequality the essential role was played by notion "a relevant (actual) state". Once again we shall dwell upon this notion. Earlier it is asserted that the nonlocal structure of a quantum object (physical state) defines a nonlinear functional~$\vp(\;)$. However, it is only a potential definition. The concrete value of the functional~$\vp(\;)$ for this or that observable~$\A$ is defined as a result of physically admissible measurement. 

The set $\{\vp \}^{\A}$ of relevant states is the range of measurability of the observable~$\A$. If the observable~$\A$ was represented by an operator on physical states, it would have a range of definition. The range of measurability is some replacement of the range of definition in that case, when the observable is not represented in the form of an operator.

We do not know a constructive scheme of composition of all relevant states at present. However, two important statements about the range of relevant states can be formulated.

1. All results of the experiments, carried out till now, are values of functionals~$\vp(\;)$ which correspond to the relevant states.

2. All results, which can be obtained in the future experiments, also correspond to the relevant states.

For deductions, which are obtained in the present work, it is sufficient that the relevant states exist, and possess these two properties.

If the algebra \AAA{} were commutative, it would be possible to consider each physical state as a relevant one. However, since the sets of relevant states for noncommuting observables cannot intersect, the set of all physical states should be divided into subsets of relevant states for various observables. This partition is multivalued. It is determined by the variant, which is chosen for the expansion of the initial definition of the sets $[\vp(\;)]^{\A}$.

By virtue of the postulate (3.4) relevant states play an essential role in definition of the linear functional on algebra~\AAA, corresponding to a quantum state. In turn, this linear functional defines a particular representation of algebra~\AAA{} by linear operators. Thus, each partition of the set of physical states fixes a particular representation of algebra~\AAA. However, the standart nonrelativistic quantum mechanics takes into account only a finite number of degrees of freedom. In this case all representations are unitary equivalent.

Summing up, it is possible to state that the new notion, introduced in the present work, "a physical state", on the one hand, does not violate the deductions of the standart quantum mechanics, and, on the other hand, allows to give a quite objective character to the principal physics of quantum mechanics. The observer occupies a place, appropriate to him. His role is essential in the description of physical events, in particular, the use of this or that quantum state depends on him. However, physical phenomena have objective character.

There are no paradoxes of type "the Schr\"odinger cat" or "the Wigner friends", Einstein-Podolsky-Rosen paradox, when a physical state is used. Of course, it is possible to adhere to the point of view that in the orthodox approach these paradoxes also are absent, if the problems are stated properly. However, it seems (see also \cc{hom}, \cc{nam}) that the answers are given not exactly to those questions, which are put by the authors of the paradoxes.

I shall also note that the use of a physical state allows to give a quite obvious physical interpretation to the process of quantum measurement and the collapse of a quantum state (see \cc{slav}).

\end{document}